\documentclass[%
journal,%
twoside,%
final,%
]{IEEEtran}%
\usepackage[T1]{fontenc}%
\usepackage{microtype}%
\usepackage{amsmath}%
\usepackage{amssymb}%
\usepackage{textcomp}%
\usepackage{mathtools}%
\usepackage{cite}%
\usepackage{xcolor}%
\usepackage{graphicx}%
\usepackage{hyperref}
\newtheorem{thm}{Theorem}%
\newtheorem{lem}{Lemma}%
\newtheorem{assum}{Assumption}%
\newtheorem{rem}{Remark}%
\newtheorem{prob}{Problem}%
\newcommand{\ie}{i\/.\/e\/.,\/~}%
\newcommand{\eg}{e\/.\/g\/.,\/~}%
\newcommand{\cf}{cf\/.\/~}%
\newcommand{\abvFig}{Fig\/.\/\,}%
\newcommand{\abvThm}{Thm\/.\/\,}%
\newcommand{\abvSec}{Sec\/.\/\,}%
\newcommand{\hoeffdingL}{L}%
\newcommand{\T}{\ensuremath{\top}}%
\DeclareMathOperator{\trace}{trace}%
\DeclareMathOperator{\proba}{\mathbb{P}}%
\DeclareMathOperator{\distributionNormal}{\mathcal{N}}%
\DeclareMathOperator{\expectation}{\mathbb{E}}%
\DeclareMathOperator{\identity}{\mathrm{I}}%
\DeclareMathOperator{\diag}{diag}%
\makeatletter
\@ifpackageloaded{hyperref}{
	\colorlet{lcolor}{blue!40!black}
	\colorlet{ucolor}{blue!50!cyan!50!black}
	\colorlet{ccolor}{green!40!black}
	\colorlet{fcolor}{red!40!black}
	\usepackage{hypcap}
	\hypersetup{colorlinks=true,linkcolor=lcolor,urlcolor=ucolor,citecolor=ccolor,filecolor=fcolor}
}{
	\providecommand{\href}{2}{#2}
	\providecommand{\orcid}{1}{}
}
\makeatother
\newcommand{\orcid}[1]{\href{https://orcid.org/#1}{\raisebox{0.2ex}{\,\includegraphics[height=1.75ex]{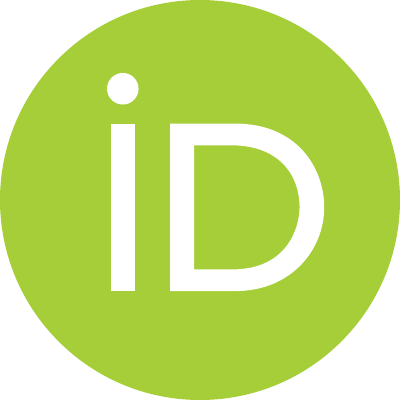}}}}
\begin{document}
	\title{Event-triggered Learning \\ for Linear Quadratic Control} 
	
	\author{
		Henning Schl\"uter$^{1,2,*}$\orcid{0000-0001-6057-2404}, %
		Friedrich Solowjow$^{1,*}$\orcid{0000-0003-2623-5652}, %
		Sebastian Trimpe$^{1,3}$\orcid{0000-0002-2785-2487}%
		\thanks{*Equally contributing.}
		\thanks{$^{1}$Intelligent Control Systems Group, Max Planck Inst. for Intelligent Systems, 70569 Stuttgart, Germany. E-mail: solowjow@is.mpg.de.}%
		\thanks{$^{2}$Institute for Systems Theory and Automatic Control, University of Stuttgart, 70550 Stuttgart, Germany. E-mail: henning.schlueter@ist.uni-stuttgart.de}
		\thanks{$^{3}$Institute for Data Science in Mechanical Engineering, RWTH Aachen University, 52068 Aachen, Germany. E-mail: trimpe@dsme.rwth-aachen.de}
		\thanks{This work was supported in part by the Cyber Valley Initiative, the International Max Planck Research School for Intelligent Systems (IMPRS-IS), and the Max Planck Society.}%
	}
	\markboth{Schl\"uter \MakeLowercase{\textit{et al.}}: Event-triggered Learning for Linear Quadratic Control}%
	{Schl\"uter \MakeLowercase{\textit{et al.}}: Event-triggered Learning for Linear Quadratic Control}
	\IEEEpubid{\parbox{\textwidth}{~\\[3\baselineskip]
		\textbf{Accepted final version.} Accepted for publication in: IEEE Transactions on Automatic Control, 2021.\\[0.1\baselineskip]		
		\textcopyright 2020 IEEE.  Personal use of this material is permitted.  Permission from IEEE must be obtained for all other uses, in any current or future media, including reprinting/republishing this material for advertising or promotional purposes, creating new collective works, for resale or redistribution to servers or lists, or reuse of any copyrighted component of this work in other works.
	}}

	\maketitle
	\begin{abstract}
		When models are inaccurate, the performance of model-based control will degrade.
		For linear quadratic control, an event-triggered learning framework is proposed that automatically detects inaccurate models and triggers the learning of a new process model when needed.
		This is achieved by analyzing the probability distribution of the linear quadratic cost and designing a learning trigger that leverages Chernoff bounds.
		In particular, whenever empirically observed cost signals are located outside the derived confidence intervals, we can provably guarantee that this is with high probability due to a model mismatch.
		With the aid of numerical and hardware experiments, we demonstrate that the proposed bounds are tight and that the event-triggered learning algorithm effectively distinguishes between inaccurate models and probabilistic effects such as process noise. Thus, a structured approach is obtained that decides when model learning is beneficial.
	\end{abstract}
	\IEEEpeerreviewmaketitle

	\section{Introduction}
	\IEEEPARstart{L}{inear} quadratic regulator (LQR) problems are well understood in literature and yield tractable and well-behaved solutions (see, for example,~\cite{aastrom2012introduction, anderson2007optimal} and references therein).
	Because of this, they are frequently used in practice, and even applications to nonlinear problems are possible with the aid of iterative methods that linearize the system dynamics~\cite{todorov2005generalized}.
	While LQR has favorable robustness properties~\cite{Lee2012}, the performance of the controller naturally depends on the accuracy of the underlying model. Thus, just like any other model-based design, LQR will generally benefit from a precise model, both in terms of performance and robustness.

\begin{figure}[tb]
		\centering
		\includegraphics[width=\linewidth]{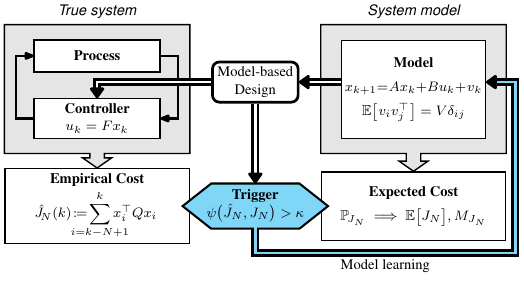}
		\caption{
			Proposed event-trigger learning architecture.
			The classic model-based controller architecture is extended by introducing the triggering of model learning when needed (blue) in order to improve performance.
			Learning experiments are triggered whenever there is a significant difference between the empirical cost, which is observed from data, and the theoretically expected cost, which is analytically derived from the model.
			After identifying an improved model, a new controller and trigger are derived based on said updated model.
			Thus, the model-based controller is closer to the underlying dynamics and, therefore, yields a reduced cost.
		}
		\label{fig:architecture}
	\end{figure}

	We propose to improve the model during operation from data \emph{when needed}. Clearly, the idea of data-driven model updates is not new~\cite{hou2013}, however, principled decision making on when to learn is a novel approach.
	Learning permanently can be wasteful from a resource point of view and may suffer from divergence issues when the system is standing still, and there is no persistent excitation in the data~\cite{aastrom2013adaptive,anderson2005failureadaptive}.
	Hence, we propose to separate the process of learning from the nominal behavior of the system and investigate the question of \emph{when to learn}.
	By automatically detecting the instances where learning is beneficial, we maintain the advantages of both data-driven and optimal control approaches by performing learning in a controlled environment and afterward, applying the rich optimal control framework to learned models. However, the crucial difficulty lies in deciding \emph{when to learn}, which we address herein with the aid of an event-triggered learning (ETL) approach, whose architecture is depicted in \abvFig \ref{fig:architecture}.

	\IEEEpubidadjcol 

	\subsection*{Related Work}
	The idea of event-triggered learning, \ie triggering model learning when needed, was first proposed in~\cite{Solowjow2018,Solowjow2019} to achieve communication savings in networked control systems, where model-based predictions are used to anticipate other agents' behavior and thus, avoiding continuous communication. Recently, the idea of event-triggered learning has been applied to event-triggered pulse control~\cite{baumann2019event} and to
	track human running gaits in sensor networks~\cite{beuchert19,beuchert20}, where it yields significant communication savings.
	In all of the above articles, the times between two communication instances are analyzed and their statistical properties are used to trigger learning updates.
	By deriving a model-induced probability distribution and simultaneously testing how likely it is that observed inter-communication times are actually drawn from this probability distribution, it is possible to construct learning triggers. Further, statistical guarantees can be deduced based on concentration inequalities, such as Hoeffding's inequality.
	In this article, we build on the main idea of ETL, but we develop learning triggers for a very different setting: models will be used for control design (instead of predictions), and triggers will be based on control performance (rather than communication). 
	In addition to the conceptual novelty, the herein proposed approach is also very different from a technical point of view. In particular, we consider Chernoff inequalities to incorporate auto-correlations in a distribution-based learning trigger.
	This is the first work to develop the idea of ETL for control. Concretely and to the best of our knowledge, this is the first approach that online monitors the performance of an LQR and triggers model learning when needed to improve performance.

	The LQR framework is a popular control method with many recent applications (see for instance~\cite{marco_ICRA_2016, trimpeCSM12, mason_full_2014, borrelli2008distributed}). Usually, the expected value of the cost function is minimized, however, there are also approaches that consider variance such as minimum variance control (see~\cite{aastrom2012introduction} for details). Closed-form solutions for the variance of the cost were derived in~\cite{Bijl2016} very recently.
	Taking this approach one step further, we consider the full distribution of the cost functional. In particular, we characterize the distribution in terms of the moment generating function. To our knowledge, this is the first such characterization of the LQR cost.
	Further, we develop learning triggers that perform goodness of fit tests that are closely designed to the problem at hand: model-based statistical properties of the cost are compared to observed empirical data. In particular, we leverage Chernoff bounds to derive confidence intervals that contain a predefined portion of the probability mass. Learning experiments are triggered whenever the empirical cost is not contained within these bounds.
	Further, we show that it is not sufficient to analyze the mean and higher moments, since there are many inherent challenges, such as auto-correlations and unbounded control costs.

	Adaptive control (see~\cite{aastrom2013adaptive, ioannou2012robust, bradtke1994adaptive} and references therein) seeks to continuously update system parameters or controllers in order to cope with changing environments. Updating the parameters permanently makes adaptive control algorithm potentially fast and flexible, however, the convergence of such algorithms is usually tightly connected to persistently excited signal vectors~\cite{bradtke1994adaptive}, which is not necessarily satisfied.
	Further, it is well known in statistical literature that simultaneous parameter estimation and testing might lead to distorted test statistics and different asymptotic distributions~\cite{kac1955tests} of statistical tests. There exist statistical tests that explicitly take the distribution of the estimator into account~\cite{babu2004goodness}, however, the dependency is often highly non-trivial.
	In our approach, we propose to separate control from learning. Furthermore, learning is only triggered when there is a significant difference to the expected behavior and hence, a difference in the signal we ultimately care about.
	Thus, we only update models and controllers when needed, which is conceptually very different from adaptive control.

	Robust control (\cf~\cite{zhou1998robust} and references therein) is also related to the proposed method, but has a different objective. The goal of robust control is designing control policies, which work decently for a variety of system parameters without changing the controller.
	In the event-triggered learning approach, we keep the controller fixed as long as the system parameters are not changing significantly. However, when there is a significant change in the system, we propose to update the model automatically.
	Thus, the proposed method possess enough flexibility to adapt to new environments, while still being efficient and robust, in particular during times with no changes in the system.

	Augmenting control architectures with machine learning techniques is a popular research direction. There are many advances such as in reinforcement learning~\cite{recht2019tour,lee2018primal}, networked control~\cite{yoo2019event, eisen2018learning,gatsis2018sample}, or model predictive control~\cite{chen2018approximating,nubert2020safe}.
	There is also recent work from a statistical point of view on LQR and sample complexity~\cite{dean2019sample, krauth2019finite, mania2019certainty} that analyzes the convergence speed of popular methods.
	However, most work revolves around learning dynamics models and data-driven improvements of controllers. In contrast to this, our work investigates the question of \emph{when to learn}, derives learning triggers, and provides theoretical guarantees for these triggers.

	The key contribution of this article lies in designing dedicated learning triggers that compare empirical costs to a model-induced distribution.
	Change and fault detection~\cite{isermann1984process, isermann2006fault, looze1985automatic, zhan1999interacting} have already been addressed in the literature, and there are many methods that can be applied to the considered problem.
	This also involves closed-loop performance assessment and monitoring \cite{qin1998control,huang1997good,swanda1999controller,dong2018novel}, which is conceptually related to our work, however, usually analyzes minimum variance control and does not consider more complex statistical objects such as moment generating functions.
	Our presented approach is specifically designed for linear Gaussian systems and the signal we care about, which is the control cost. Thus, we obtain an efficient algorithm with tight confidence bounds that are based on the herein derived expression for the moment generating function of the cost.

	\subsection*{Contributions}
	To summarize, this article makes the following contributions:
	\begin{itemize}
		\item introducing the concept of control performance-based event-triggered learning for linear Gaussian systems -- the model and therefrom derived quantities are only updated when there are significant changes in the online performance;
		\item characterization of the full distribution of the LQR cost functional via moment generating functions (\abvThm\ref{thm:mgf});
		\item design of an effective learning trigger based on the full distribution (\abvSec\ref{sec:chernoff}) of the LQR cost with additional theoretical guarantees, which are derived with the aid of concentration inequalities and demonstration that considering moments (\ie the expected value) is not sufficient; and
		\item validation and comparison of the derived triggers in simulation and hardware experiments, in which we demonstrate fast, reliable, and robust detection.
	\end{itemize}

	\subsection*{Notation}
	The maximum eigenvalue of a matrix $Q$ is denoted $\lambda_{\mathrm{max}}(Q)$.
	For a positive definite matrix $Q$, we denote principal square root $\sqrt{Q}$, \ie the positive definite matrix satisfying $\sqrt{Q}\sqrt{Q}=Q$.
	The Kronecker delta $\delta_{ij}$ is equal to one if $i=j$ and zero otherwise.
	We use $\proba[s\in\mathbb{S}]$, $\proba[s\in\mathbb{S},t\in\mathbb{T}]$, and the notation $\proba[s\in\mathbb{S}|t\in\mathbb{T}]$ for the probability of an event $s\in\mathbb{S}$, the joint probability of the events $s\in\mathbb{S}$ and $t\in\mathbb{T}$, and the conditional probability of event $s\in\mathbb{S}$ given $t\in\mathbb{T}$.
	Further, we denote the respective probability density functions of the random variables $s$ and $t$ with $\proba[s]$, $\proba[s,t]$, and $\proba[s|t]$ .
	The expected value of a random variable $s$ is $\expectation[s]$.

	\section{Event-Triggered Learning for LQR control}
	In this section, we formulate the problem of event-triggered learning for linear quadratic control and present the main ideas of this article.

	\subsection{Problem setup}
	We assume the linear dynamics
	\begin{equation}
	x_{k+1} = A_\mathrm{o} x_k + B u_k + v_k, \label{eq:system:open-loop}
	\end{equation}
	with discrete-time index $k\in \mathbb{N}$, state $x_k \in \mathbb{R}^n$, control input $u_k \in \mathbb{R}^q$, system matrix $A_\mathrm{o} \in \mathbb{R}^{n \times n}$, input matrix $B \in \mathbb{R}^{n \times q}$ and independent identically distributed (i.i.d.) Gaussian noise $v_k \sim \distributionNormal(0,V)$ with $\expectation\!\left[v_iv_j^\T\right]=V\delta_{ij}$.
	Further, the system is assumed to be $(A_\mathrm{o},B)$-stabilizable.
	Hence, stable closed-loop dynamics can be achieved through state feedback
	\begin{equation}
	u_k = -F x_k + u_\mathrm{ref}(k), \label{eq:feedback_law}
	\end{equation}
	where $F \in \mathbb{R}^{q \times n}$ is the feedback gain and $u_\mathrm{ref}(k)$ is a known reference, which can be used to track a trajectory or excite the system in order to generate informative data.

	A stabilizing feedback gain can be obtained, for instance, via LQR design~\cite{anderson2007optimal}.
	In particular, we can use Riccati equations to find analytical solutions to the optimal control problem with the quadratic cost function
	\begin{equation}
	J = \lim_{N\to\infty}\frac{1}{N}\,\mathbb{E}\!\left[\sum_{j=0}^{N-1} x_j^\T Q_\mathrm{LQR} x_j + u_j^\T R_\mathrm{LQR} u_j\right],
	\end{equation}
	where $Q_\mathrm{LQR}$ and $R_\mathrm{LQR}$ are symmetric and positive definite matrices with compatible dimensions.
	In the following, we consider the empirical cost over a finite horizon $N$, which we will denote at time step $k$ by
	\begin{equation}
	\hat{J}_N(k) = \sum_{\mathclap{j=k-N+1}}^{k} x_j^\T Q_\mathrm{LQR} x_j + u_j^\T R_\mathrm{LQR} u_j\,.
	\end{equation}
	A normalization is not needed here since the cost will remain finite when considering a finite horizon.
	Thus, we will drop the normalization for notational convenience since it has no theoretical influence on the later obtained results.

	To further ease the notation, we write
	\begin{equation}
	x_{k+1} = A x_k + v_k \label{eq:sys},
	\end{equation}
	with $A=(A_\mathrm{o}-BF)$ and obtain
	\begin{equation}
		\hat{J}_N(k) = \sum_{\mathclap{j=k-N+1}}^{k} x_j^\T Q x_j \label{eq:cost},
	\end{equation}
	where $Q=\left(Q_\mathrm{LQR} + F^\T R_\mathrm{LQR} F\right)$.

	It is well-known that the states of a stable system (such as \eqref{eq:sys}) converge to a stationary Gaussian distribution.
	In particular, the steady-state covariance $X^{V} \coloneqq \lim_{k\rightarrow\infty}\expectation[x_kx_k^\T]$ can be computed as the solution to the Lyapunov equation (\eg~\cite[Lemma~2.1]{Kuvaritakis2014})
	\begin{equation}
		A X^{V}\!\! A^\T - X^{V} + V = 0.
	\end{equation}

	The stationary state covariance $X^{V}$ is a key object for the following technical development, and thus, we want to explicitly point out the technical assumptions that are necessary.
	\begin{assum}\label{asm:stability}
		The closed-loop model \eqref{eq:sys} is stable in the sense that $|\lambda_{\mathrm{max}}(A)| < 1$.
	\end{assum}

	This assumption is not very restrictive, as we only require the feedback law \eqref{eq:feedback_law} to stabilize the open-loop model \eqref{eq:system:open-loop}.
	\begin{assum}\label{asm:convergence}
		The system has converged to a steady state, in the sense that $\expectation\!\left[x_k\right] = 0$ and the covariance $\expectation[x_kx_k^\T] = X^{V}$ are time-invariant.
	\end{assum}

	Given Assumption~\ref{asm:stability}, it follows directly that the system converges exponentially to a steady-state Gaussian distribution~\cite[\abvSec3.1]{Kumar1986}.
	The problem can easily be generalized to $\expectation\!\left[x_k\right]=\mu$ by subtracting the constant mean from given data.
	Thus, the assumptions made here are not very strong.

	In the following, we distinguish between the model-induced cost $J_N$, which is a random variable, and the empirical cost $\hat{J}_N(k)$, which is sampled from the system.
	For the random variable $J_N$, we can drop the dependency on $k$. This follows directly from assuming stationary states in Assumption~\ref{asm:convergence}. Since we are considering quadratic transformations of stationary random variables (the states) and the summation over a fixed window of length $N$, the random variable $J_N$ is itself stationary. Thus, we can omit the index $k$.

	\subsection{Problem and Main Idea}
	In this work, we systematically analyze the question of \emph{when to learn} a new model of the dynamical system \eqref{eq:system:open-loop}, which is later on utilized to synthesize a controller.
	Due to the structure of the problem, we are able to quantify how well the controller should perform in terms of expected value, variance, or a distributional sense.
	The statistical testing is carried out under the null hypothesis that model and ground truth coincide.
	Thus, by checking if theoretically derived properties actually coincide with empirically observed cost values, we are able to detect significant mismatches between the current model and the ground truth dynamics.

	This idea leads to the proposed ETL architecture shown in \abvFig\ref{fig:architecture}.
	The core piece of the proposed method is the binary event trigger $\gamma_\mathrm{learn}$ for learning a new model and the corresponding test statistic $\psi$ that quantifies how likely it is that empirical samples $\hat{J}_N(k)$ coincide with the model-induced random variable $J_N$. Given a level of confidence $\eta$, we are able to compute critical thresholds $\kappa$ and, thus, trigger learning experiments on necessity. Since we are considering linear systems here, the main emphasis is on the design of the test statistic $\psi$. Identifying linear systems is not the focus of this article and has been extensively discussed in previous work (see~\cite{ljung1998system} for an overview). After a new model is identified, we propose to compute a new controller and derive new trigger thresholds. We thus summarize the core problem addressed in this article.

 	\begin{prob}
	 	Detect, when the model has changed, by comparing the deviation of model-induced cost properties to empirical costs, thus, yielding the learning trigger
		\begin{equation}
		\psi\bigl(\hat{J}_N, J_N\bigr) > \kappa \Leftrightarrow \gamma_\mathrm{learn} = 1,
		\end{equation}
		where $\psi$ is an appropriate test statistic, $\kappa$ is the computed critical threshold and $\gamma_\mathrm{learn}$ is a binary indicator for whether a model update is required ($\gamma_\mathrm{learn}=1$) or not ($\gamma_\mathrm{learn}=0$).
	\end{prob}

	Due to the Gaussian process noise, the proposed trigger will also exhibit an expected probabilistic behavior.
	In particular, it is impossible to entirely avoid false positive learning decisions. Therefore, we take this explicitly into account when designing the learning trigger by choosing $\kappa$ such that
	\begin{equation}
		\proba\!\left[\psi\bigl(\hat{J}_N, J_N\bigr) > \kappa\right]<\eta\,,
	\end{equation}
	\ie the probability of the trigger misfiring is less then desired confidence level $\eta$.

	First, we develop a learning trigger incorporating the entire distribution in the form of a moment generating function in \abvSec\ref{sec:chernoff}, which allows for an efficient implementation to detect system changes.
	Then, we demonstrate that the straightforward approaches using single moments of the cost (\abvSec\ref{sec:hoeffding}) pose theoretical challenges. 
	For the case of the expected value, we show that the trigger based on moment generating function with its superior theoretical properties also yields better empirical performance and reliability.

	\section{Distribution-based Trigger}\label{sec:chernoff}
	In this section, we derive a learning trigger, which is directly based on 
	confidence bounds of the empirical cost.
	The idea is to apply the Chernoff bound to the empirical cost in order to obtain a likely range of values.
	Then, the trigger can easily detect values outside of this interval.
	Before we can apply the Chernoff bound, we first need to derive the moment-generating function (MGF) of the cost function.

	\subsection{Moment Generating Functions}\label{sec:chernoff:mgf}
	The moment-generating function (MGF) $M_X(\xi)\coloneqq\expectation[e^{\xi X}]$ of a random variable $X$ -- if it exists -- is a powerful tool to characterize the distribution (see, \eg~\cite[Chapter 4]{gut2013probability} for more details on MGFs).
	It is moment-generating in the sense that for all $n\in \mathbb{N}$, the $n$-th moment $\expectation[X^n]$ can be obtained by computing $\frac{\mathrm{d}}{\mathrm{d}\xi} M_{J_N}\!(\xi)|_{\xi=0}$.

	Next, we will compute the MGF of the cost $J_N$ and afterward, combine it with Chernoff bounds to obtain a powerful trigger.
	\begin{thm}[Moment Generating Function of the Cost]\label{thm:mgf}
		Assuming the state sequence $z=\left(x_0, x_1, \dotsc, x_{N-1}\right)^\T$ is a jointly normally distributed random variable with mean $\mu$ and covariance matrix $\Sigma$.
		The moment generating function of the cost $J_N=z^\T\Omega z=\sum_{k=0}^{N-1}x_k^\T Q x_k$ is given by
		\begin{equation}
		M_{J_N}\!\left(\xi;\mu,\Sigma,\Omega\right)
		= \frac{\exp\!\left(\frac{1}{2}\mu^\T\!\left(\!\left(\identity - 2 \xi \Omega\Sigma\right)^{-1} - \identity\right)\!\Sigma^{-1}\mu \right)}{\sqrt{\det\!\left(\identity - 2\xi\Omega\Sigma\right)}},
		\end{equation}
		where $\xi \in \bigl[-\infty, \tfrac{1}{2 \lambda_{\max}\!\left(\Omega\Sigma\right)}\bigr)$ and $\Omega = \diag\!\left(Q,  \dotsc, Q\right)$ with weight matrix $Q$.
	\end{thm}
	\begin{IEEEproof}
		It is a known fact that there exits an $m\times m$ matrix $T$ such that $\det T\ne 0$, $T^\T \Sigma^{-1} T = \identity$, $T^\T\Omega T = \Lambda$, and $\Sigma\Omega=T^{-\T}\Lambda T^\T$, where $\Lambda$ has the eigenvalues $\lambda_i$ of $\Sigma\Omega$ on the diagonal, given that $\Sigma$ and $\Omega$ are symmetric and $\Sigma$ is positive definite.
		As both $\Sigma$ and $\Omega$ fulfill this requirement by definition, we can use this to obtain a transformation matrix $T$.
		Let $F_z$ denote the cumulative distribution function of $z$, \ie of a normal distribution.
		It then follows by definition that
		\begin{align*}
		&M_{J_N}(\xi)= \expectation\left[e^{\xi z^\T\Omega z}\right] = \int_{\mathbb{R}^{Nn}} \exp \!\left(\xi x^\T \Omega x\right)\!\mathrm{d}F_z(x)\\
		&= \det\left(2\pi\Sigma\right)^{-\frac{1}{2}} \!\!\int\! \exp \!\left(\xi x^\T \Omega x - \tfrac{1}{2} \left(x \!-\! \mu\right)^\T \!\Sigma^{-1}\! \left(x \!-\! \mu \right) \right)\!\mathrm{d}x,
		\end{align*}
		where $\int\mathrm{d}x$ is an $m$-fold integral over the domain of $z$, \ie $\mathbb{R}^{Nn}$.
		Applying the transformation $x=Ty+\mu$ with $c=T^{-1}\mu=(c_1,\dotsc,c_m)$, we rewrite as
		\begin{align*}
		M_{J_N}(\xi) &= \prod_{i=1}^{m}\frac{1}{\sqrt{2\pi}}\int_{-\infty}^{\infty}\exp\left(\xi\lambda_i(y_i+c_i)^2-\frac{1}{2}y_i^2\right)\!\mathrm{d}y_i\\
		&=\left[\,\prod_{i=1}^{m}\frac{1}{1-2\xi\lambda_i}\right]\cdot\left[\exp\sum_{i=1}^{m}\frac{1}{2}c_i^2\frac{2\xi\lambda_i}{1-2\xi\lambda_i}\right]\\
		&=\frac{\exp\!\left(\frac{1}{2}\mu^\T\!\left(\!\left(\identity - 2 \xi \Omega\Sigma\right)^{-1} - \identity\right)\!\Sigma^{-1}\mu \right)}{\sqrt{\det\!\left(\identity - 2\xi\Omega\Sigma\right)}}
		\end{align*}
		\nopagebreak\vspace{-1.75\baselineskip}\\
	\end{IEEEproof}

	By Assumption~\ref{asm:stability} and \ref{asm:convergence}, we have $\mu=0$ and $\Sigma=\text{const}$.
	This yields	the simplified form
	\begin{equation}
	M_{J_N}\!\left(\xi;\Sigma,\Omega\right)=\det\!\left(\identity - 2\xi\Omega\Sigma\right)^{-\frac{1}{2}}, \label{eq:mgf:simple}
	\end{equation}
	which is time-invariant as it only depends on constant model parameters.
	Thus, it is also well suited for the design of a learning trigger.

    Based on the MGF, it is straightforward to compute the moments.
	\begin{lem}\label{lem:mgf:moments}
		For $\expectation\!\left[x\right] = 0$, the expected value and variance for the cost $J_N$ of a trajectory $x$, as derived from the moment-generating function $M_{J_N}(\xi)$, are given by
		\begin{subequations}\begin{alignat}{3}%
		\expectation\bigl[J_N\bigr]%
		&=&\,\frac{\mathrm{d}}{\mathrm{d}\xi} M_{J_N}\!\left(\xi\right)\biggr\rvert_{\rlap{\ensuremath{\scriptstyle\xi=0}}}%
		&=\trace\Omega\Sigma\\%
		\expectation\bigl[J_N^2\bigr]%
		&=&\,\frac{\mathrm{d}^2}{\mathrm{d}\xi^2} M_{J_N}\!\left(\xi\right)\biggr\rvert_{\rlap{\ensuremath{\scriptstyle\xi=0}}}%
		&=2\trace\!\left(\Omega\Sigma\right)^2+\trace^2\Omega\Sigma\,.%
		\end{alignat}\end{subequations}
	\end{lem}
	\begin{IEEEproof}
		The result follows from using Jacobi's formula $\tfrac{\mathrm{d}}{\mathrm{d}t} \det A(t) = \det A(t) \trace \left(A^{-1} \tfrac{\mathrm{d}}{\mathrm{d}t}A(t)\right)$ ~\cite[\abvSec8.3]{Magnus1999} to compute the derivatives of the moment-generating function.
	\end{IEEEproof}

	\subsection{Chernoff Trigger}\label{sec:chernoff:trigger}
	In order to obtain an effective trigger with theoretical guarantees, we need sophisticated concentration results, which make use of the whole distribution.

	Next, we introduce the Chernoff bound and utilize it to derive the trigger threshold $\kappa$.
	\begin{thm}[Chernoff Bound \mbox{~\cite[\abvThm1]{Chernoff1952}}] \label{thm:chernoff:inequality}
		Given the moment\/-\/generating function $\expectation\!\left[\mathrm{e}^{\xi X}\right]$ of the random variable X, for any real number $\xi>0$, it holds that
		\begin{align}
		\proba\!\left[X \ge \kappa\right]\! &\le \frac{M_X(\xi)}{\mathrm{e}^{\xi \kappa}} &
		\proba\!\left[X \le \kappa\right]\! &\le \frac{M_X(-\xi)}{\mathrm{e}^{-\xi \kappa}}.\\
		\shortintertext{In particular, it holds that}
		\proba\!\left[X \ge \kappa\right]\! &\le \inf_{\xi>0} \frac{M_X(\xi)}{\mathrm{e}^{\xi \kappa}} &
		\proba\!\left[X \le \kappa\right]\! &\le \inf_{\xi<0} \frac{M_X(\xi)}{\mathrm{e}^{\xi \kappa}}.
		\end{align}
	\end{thm}
	\begin{IEEEproof}
		Follows from Markov's inequality applied to~$\mathrm{e}^{\xi X}$.
	\end{IEEEproof}
	\begin{rem}
	These bounds are often specialized to sums of independent random variables. We avoid this independence requirement, by considering the MGF of the entire sum as a random variable. This enables us to tailor the Chenoff trigger to the full distribution of $J_N$ taking any dependence over the horizon into account. This is in contrast to the moment-based trigger designed in the next section.
	\end{rem}
	
	Thus, we can state the main theorem of this article, which is the full distributional analog to \abvThm\ref{thm:hoeffding:trigger}.
	\begin{thm}[Chernoff Trigger]\label{thm:chernoff:trigger}
		Let the parameter $N\in\mathbb{N}$ be given and
		Assumptions \ref{asm:stability} and \ref{asm:convergence} hold. Then, we can obtain for any time-index $k$  an upper bound $\eta\in\left(0,1\right)$ for the probability
		\begin{equation}
		\proba\!\left[J_{N}(k)\notin\left(\kappa^-,\kappa^+\right)\right]\! \le \eta,
		\end{equation}
		where the thresholds are chosen in the following
		\begingroup\setlength{\belowdisplayskip}{0pt}\setlength{\belowdisplayshortskip}{0pt}%
		\begin{align}
		\kappa^+ &= {\inf_{\xi \in \left(\!0, \frac{1}{2 \lambda_{\max}}\!\right)}} \chi(\xi)
		&
		\kappa^- &= {\sup_{\xi \in \left(-\infty, 0\right)}} \chi(\xi)
		\label{eq:kappachernoff}
		\end{align}
		\endgroup\begingroup\setlength{\abovedisplayskip}{0pt}\setlength{\abovedisplayshortskip}{0pt}%
		\begin{subequations}\begin{align}
			\chi(\xi) &= -\tfrac{1}{\xi}\ln \tfrac{\eta}{2} - \tfrac{1}{2 \xi}  \ln\det\!\left(\identity-2\xi\Omega\Sigma\right)\\
			&= -\frac{1}{\xi}\ln \frac{\eta}{2} - \frac{1}{2 \xi} \sum_{j=0}^{Nn} \ln\!\left(1-2\xi\lambda_j\right).
		\end{align}\end{subequations}\endgroup
		Further, $\lambda_j$ are the eigenvalues of $\Omega\Sigma$, the state covariance matrix is denoted as $\Sigma$ (as introduced in Lemma \ref{lemma:invariant_joint_state_distribution}), and the weight matrix $\Omega = \diag(Q,\linebreak[0]\dotsc,\linebreak[0]Q)$.
	\end{thm}
	\begin{IEEEproof}
		We distribute the tail probability $\eta$ symmetrically to both sides of the interval.
		Thus,
		\begin{equation*}
			\inf_{\xi>0} \frac{M_X(\xi)}{\mathrm{e}^{\xi \kappa^+}} = \inf_{\xi<0} \frac{M_X(\xi)}{\mathrm{e}^{\xi \kappa^-}} = \frac{\eta}{2},
		\end{equation*}
		which has to be solved for $\kappa^\pm$.
		For $\kappa^+$, we get
		\begin{alignat*}{3}
			&\qquad & \frac{\eta}{2} &= \inf_{\xi>0} \frac{M_X(\xi)}{\mathrm{e}^{\xi \kappa^+}}\\
			&\Leftrightarrow & 0 &= \inf_{\xi>0} \ln{M_X(\xi)}-{\xi \kappa^+}-\ln\tfrac{\eta}{2} \mathrlap{\smash{\quad \bigg|\div \xi}}\\
			&\Leftrightarrow & \kappa^+ &= \inf_{\xi>0} \tfrac{1}{\xi}\ln{M_X(\xi)}-\tfrac{1}{\xi}\ln\tfrac{\eta}{2}
		\end{alignat*}
		We can proceed similarly for $\kappa^-$, just that the infimum turns into a supremum, when we divide by $\xi$ as $\xi<0$.
		By inserting the simplified MGF from \eqref{eq:mgf:simple} into the equation, we obtain the statement.
	\end{IEEEproof}
	
	\begin{figure}
		\centering
		\includegraphics[width=\linewidth]{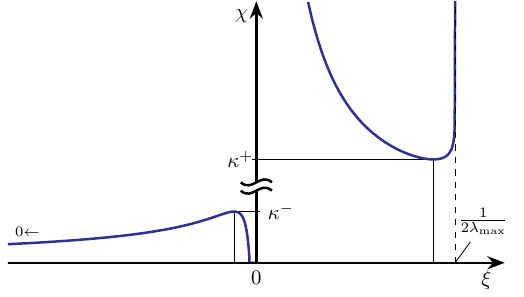}
		\caption{Illustration of the shape of the function $\chi(\xi)$, which has to be optimized for the Chernoff trigger. The scales of the left- and right-hand side of the graph differ and have been adjusted for better visualization. The bounds of the Chernoff trigger, $\kappa^-$ and $\kappa^+$, are found through straightforward maximization (over $\xi$ < 0) and minimization ($\xi$ > 0), respectively. }
		\label{fig:chernoff:optfun}
	\end{figure}

	Next, we introduce the trigger design, discuss the main advantages, and, finally, elaborate on how to obtain the thresholds $\kappa^\pm$.

	The Chernoff trigger is defined as follows:
	\begin{equation}\label{eq:chernoff:trigger}
	   \hat{J}_N(j) \notin \left(\kappa^-,\kappa^+\right) \Leftrightarrow \gamma_\text{learn} = 1,
	\end{equation}
	with $\kappa^\pm$ as introduced in \abvThm\ref{thm:chernoff:trigger}.

	In order to obtain the thresholds $\kappa^\pm$, we need to solve the two optimization problems \eqref{eq:kappachernoff}. However, this is easily tractable online due to the following properties of the objective function $\chi$. Intuitively this can also be seen in \abvFig\ref{fig:chernoff:optfun}, where the general shape of the objective function is illustrated.
	\begin{thm}
		The function $\chi(\xi)$ is strictly convex in the range for $\kappa^+$ and thus has only one minimum on the interval.
	\end{thm}
	\begin{IEEEproof}
		We first consider the strict convexity on the interval $0<\xi<\frac{1}{2 \lambda_{\max}}$.
		For all $\xi$ from that interval, $\frac{1}{\xi}$ is convex, thus also $-\frac{1}{\xi}\ln\frac{\eta}{2}$ is convex as $\eta\in(0,1)$ implies that the logarithm is negative.
		The second derivative of the second part is
		\begin{multline*}
		\frac{\mathrm{d}^2}{\mathrm{d}\xi^2}\biggr[-\frac{\ln\!\left(1-2\xi\lambda_j\right)}{2 \xi} \biggr]
		=
		-\frac{\ln\!\left(1-2\xi\lambda_j\right)}{\xi^3}\\
		+\frac{2\lambda_j}{ \xi^2\left(1-2\xi\lambda_j\right)}
		+\frac{4\lambda_j^2}{2\xi\left(1-2\xi\lambda_j\right)^2}.
		\end{multline*}
		In the considered interval, we have $\xi>0$, $0<1-2\xi\lambda_j<1$ and $\ln(1-2\xi\lambda_j)<0$,
		therefore, the second derivative is positive in this range.
		Hence, all summands of $\chi(\xi)$ are strictly convex. Thus, $\chi(\xi)$ is strictly convex on the interval.
		This immediately implies that there is only one minimum on the interval.
	\end{IEEEproof}

	\begin{rem}
		Even if the optimization does not yield the optimal value, any sub-optimal value will still fulfill the Chernoff bound. Thus the trigger remains valid, just with a more conservative threshold.
	\end{rem}

	\section{Mean-based Learning Trigger}\label{sec:hoeffding}
	The previous design considers only a single sample of the cost function. Intuitively one may want to use moving averages and higher moments of the cost since it is a stochastic process. Similar ideas, which are based on the expected values and  Hoeffding's inequality were presented
	in~\cite{Solowjow2019,gatsis2018sample,baumann2019event}.
	While conceptually simpler, we illustrate that moment-based triggers involve theoretical obstacles and limitations arising from this concept.
	
	The idea for this trigger is to derive a threshold $\kappa$ on the deviation from the expected value $\expectation[J_N]$, leading to
	\begin{equation}
		\label{eq:hoeffding:trigger}
		\left|\sum\nolimits_{j\in\mathfrak{\hoeffdingL}(k)} \hat{J}_N(j) - {\hoeffdingL}\expectation[J_N]\right| \ge \kappa \Leftrightarrow \gamma_\text{learn} = 1\,,
	\end{equation}
	where $\mathfrak{\hoeffdingL}(k)$ is a summation set of cardinality ${\hoeffdingL}$ that achieves approximately uncorrelated samples and will be discussed later (\cf \eqref{eq:hoeffding:summation_set}). We will first provide a derivation of the expected value, then discuss how to design the threshold $\kappa$ in order to obtain a confidence interval corresponding to a given probability.

	\subsection{Expected Value}
	Analogously to the continuous-time solution provided in~\cite{Bijl2016}, we will next derive the expected value of the cost $\expectation\!\left[J_N\right]$ for the discrete-time case.
	\begin{lem}[Expected Cost]\label{lem:expected_cost}
		Under Assumption~\ref{asm:stability} and~\ref{asm:convergence}, the expected value for the cost $J_N$ 
		with respect to the system \eqref{eq:sys} is given by
		\begin{align}
		\expectation\!\left[J_N\right]
		&= \expectation\!\left[\sum_{k=0}^{N-1} x_k^\T Q x_k\right] =\trace\left(\sum_{k=0}^{N-1} S_k Q\right)\notag\\
		&= \trace\left(\left(S_0-S_{N}+N V\right)\bar{X}^{Q}\right)\notag\\
		&= \trace\left(N V\bar{X}^{Q}\right)\,,
		\end{align}
		with $S_k=\expectation\!\left[x_kx_k^\T\right]$, and $\bar{X}^{Q}$ the solution to the Lyapunov equation
		$A^\T \bar{X}^{Q} A - \bar{X}^{Q} + Q = 0$.
	\end{lem}
	\begin{IEEEproof}
		We first note that
		$\expectation\!\left[J_N\right] = \expectation\!\left[\sum_{k=0}^{N-1} x_k^\T Q x_k\right]
		= \expectation\!\left[\sum_{k=0}^{N-1} \trace\left(x_k x_k^\T Q\right)\right]
		= \trace\left(\sum_{k=0}^{N-1} \expectation\!\left[x_k x_k^\T\right] Q\right)
		= \trace\left(\sum_{k=0}^{N-1} S_k Q\right)
		$. Then, let $Y\left(N\right)\coloneqq\sum_{k=0}^{N-1} S_k$. Next, we can find $Y(N)$ as the solution to a discrete Lyapunov equation by reordering the difference of initial and final second moment
		\begin{align*}
		S_N-S_0&=\sum_{k=0}^{N-1} \left(S_{k+1}-S_k\right) = \sum_{k=0}^{N-1} \left(A S_{k} A^\T-S_k+V \right)\notag\\
		&=A \left(\sum_{k=0}^{N-1} S_{k}\right) A^\T-\left(\sum_{k=0}^{N-1} S_{k}\right)+\sum_{k=0}^{N-1} V\notag\\
		&=A Y\!\left(N\right) A^\T - Y\!\left(N\right) + N V.
		\end{align*}
		One can show by substituting the Lyapunov equations, that $\expectation\!\left[J_N\right] = \trace(Y(N)Q)=\trace\left(\left(S_0-S_{N}+N V\right)\bar{X}^{Q}\right)$ with $Y(N)$ and $\bar{X}^{Q}$ being the solution to $0=A Y\!\left(N\right) A^\T - Y\left(N\right) +S_0-S_N+N V$ and $0=A^\T \bar{X}^{Q} A - \bar{X}^{Q} + Q$.
		With Assumption~\ref{asm:convergence} the covariance is time-invariant, thus the result simplifies to $\expectation\!\left[J_N\right] = \trace\left(N V\bar{X}^{Q}\right)$.
	\end{IEEEproof}
   	This result is equivalent to the previous result from Lemma~\ref{lem:mgf:moments} for the first moment.
   	However, here we avoid explicitly computing $\Sigma$, which is more computationally efficient for individual moments than the previous direct solution, where $\Sigma$ is explicit.

	Next, we will consider how to design the threshold $\kappa$.

	\subsection{Statistical Guarantees}
	The trigger \eqref{eq:hoeffding:trigger} leverages the fact that the empirical mean converges to the expected value. Even for finite sample sizes, it is possible to quantify the expected deviation, which can be done with the well-known Hoeffding inequality.
	The trigger threshold $\kappa$ 
	can be regarded as a confidence bound, \ie it is chosen such that with confidence level $\eta$, the deviation term does not exceed $\kappa$.
	Therefore, observing deviations larger than $\kappa$ can not be sufficiently explained by noise and random fluctuations. Thus, we trigger model learning whenever this happens.
	\begin{thm}[Hoeffding's Inequality \mbox{~\cite[\abvThm2]{Hoeffding1963}}]\label{thm:hoeffding:inequality}
		Assume $X_1,\linebreak[0] X_2,\linebreak[0] \dotsc,\linebreak[0] X_n$ are independent random variables and $a_i \le X_i \le b_i$ ($i=1,2,\dotsc,n$), then we obtain for all $\kappa>0$
		\begin{equation}
		\proba\!\left[\sum_{i=1}^n X_i-n\mu\ge \kappa\right] \le \exp\left(\frac{-2\kappa^2}{\sum_{i=1}^n \left(b_i-a_i\right)^2}\right).
		\end{equation}
	\end{thm}

	Comparing with \eqref{eq:hoeffding:trigger}, the cost samples $\hat{J}_N$ corresponds to the random variables $X_i$ and $\expectation[J_N]$ to the mean $\mu$.
	However, there are two challenges when applying Hoeffding directly in this way.
	First, $J_N$ is unbounded, as it is directly influenced by Gaussian noise.
	Second, the cost samples are not independent, as they are part of the same state trajectory.
	In the following, we introduce two modifications to cope with these issues and make Hoeffding's inequality applicable.

	In order to obtain an upper bound on $J_N$, we will assume state constraints and, for the sake of simplicity, we shall assume linear constraints.
	\begin{lem}\label{lem:hoeffding:bound}
		Assume the states are constrained by $\left\lVert W^{-1}x_k\right\rVert<\alpha$ for all $k$, where $W\in\mathbb{R}^{n\times n}$ is invertible. Then, the cost function $J_N$ is bounded by
		\begin{equation}
		0 \le J_N \le \sup_{\left\lVert W^{-1}x_k\right\rVert_2<\alpha} J_N = \alpha^2 N \lambda_{\max}\!\left(W^\T Q W\right)\,.
		\end{equation}
	\end{lem}
	\begin{IEEEproof}
		The lower bound follows immediately from the positive definiteness of $Q$, as $x^\T Q x \geq 0$ for all $x$.
		For the upper bound we use the convexity of the cost function.
		The supremum of a convex function on an open set is attained at the maximum on the boundary.
		Hence,
		\begin{align*}
		J_N &\le \sup_{\left\lVert W^{-1}x_k\right\rVert_2<\alpha} J_N
		= \max_{\left\lVert W^{-1}x_k\right\rVert_2=\alpha} \!\left[\:\sum_{k=0}^{N-1}x_k^\T Q x_k\right]\\
		&= N \max_{\left\lVert y\right\rVert_2=\alpha} \left\lVert \sqrt{Q\,}\, W y\right\rVert_2^2
		= \alpha^2 N \left\lVert \sqrt{Q\,}\, W\right\rVert_2^2\\
		&= \alpha^2 N \lambda_{\max}\!\left(W^\T Q W\right).
		\end{align*}
		\nopagebreak\vspace{-2\baselineskip}\\
	\end{IEEEproof}
	\begin{rem}
		Even for naturally unconstrained system, considering Assumptions~\ref{asm:stability} and \ref{asm:convergence}, it is reasonable to assume that the state stays within some sufficiently large, but finite, region around the origin.
	\end{rem}

	Next, we investigate how to cope with the dependence in the cost samples.
	First, we note that consecutive samples $J_N(k-1)$ and $J_N(k)$ are dependent, as they overlap in the states they sum over.
	Also, adjacent sample $J_N(k-N)$ and $J_N(k)$ are dependent, since the first state in $J_N(k)$ just follow the last state in $J_N(k-N)$.
	In order to find \emph{approximately independent} samples $J_N(j)$, we first need to consider the correlation between states in a trajectory.
	\begin{lem}\label{lemma:invariant_joint_state_distribution}
		By Assumption~\ref{asm:convergence} we have $x_0 \sim \mathcal{N}\!\left(0, X^{V}\right)$.
		Then, the joint distribution of a sequence of states $\left(x_0,x_1,\dotsc,x_N\right)$ is a multivariate Gaussian distribution with mean $\mu=0$ and symmetric block-Toeplitz covariance matrix
		\begin{gather*}
		\Sigma =\!\left(
		\begin{matrix}
		X^{V}		  & X^{V}\! A^\T & X^{V}\! (A^2)^\T & \cdots & X^{V}\! (A^{N})^\T	 \\
		A X^{V}		& X^{V}		& X^{V}\! A^\T	 & \ddots & \!X^{V}\! (A^{N-1})^\T \\
		A^2\! X^{V}	& A X^{V}	  & \ddots		 & \ddots & \vdots			   \\
		\vdots	   & \ddots	 & \ddots		 & X^{V}	& X^{V}\! A^\T		   \\
		A^N\! X^{V}\!  & \cdots	 & A^2\! X^{V}	  & A X^{V}  & X^{V}				  \\
		\end{matrix}\!\right).
		\end{gather*}
	\end{lem}
	\begin{IEEEproof}
		The covariance $X^{V}$ is invariant under the system equation, thus $\expectation[x_ix_i^\T]=X^{V}$ for all $i=0,\dotsc,N$.
		Computing the cross-covariance for two states $x_i$ and $x_j$, for $i<j$ yields $\expectation[x_ix_j^\T]=A^{i-j}X^{V}$.
		As the joint distribution over multivariate Gaussians, \ie the states, is also multivariate Gaussian, the statement follows.
	\end{IEEEproof}

	\begin{lem}\label{lem:hoeffding:independence}
		Under Assumptions~\ref{asm:stability} and~\ref{asm:convergence}, and with arbitrarily small  $\varepsilon>0$, there exist an $r_0$ such that $\bigl\lvert\left\lbrack A^{r_0}X^{V}\right\rbrack_{i,j}\bigr\rvert < \varepsilon$ for all matrix-entries $(i,j)$.
		Hence, for any large enough $r>r_0$ the state $x_k$ is approximately independent from the state $x_{k-r}$.
	\end{lem}
	\begin{IEEEproof}
		Using Lemma~\ref{lemma:invariant_joint_state_distribution}, we obtain $\expectation\!\left[x_kx_{k-r_0}^\T\right] = A^{r_0}\! X^{V}$ as the cross-covariance for the jointly multivariate normal distributed states.
		For multivariate normal distributions, we have that zero cross-covariance is equivalent to independence.
		Since, as $A$ is Schur-stable by Assumption~\ref{asm:stability}, \ie $\lambda_\mathrm{max}(A)<1$, the term $A^{r_0}$ approaches zeros as ${r_0}\to \infty$~\cite{Oldenburger1940}.
		Hence, by definition of the limit, there exists an ${r_0}$ such that the absolute value of the cross-covariance is elementwise smaller than $\varepsilon$.
		Furthermore, the same holds trivially true for any $r>r_0$.
		Therefore, the states from the same trajectory with distance $r$ are approximately independent.
	\end{IEEEproof}

	\begin{figure}
		\centering
		\includegraphics[width=\linewidth]{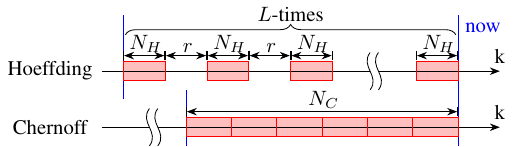}
		\caption{
			Sampling intervals for Hoeffding (\abvSec\ref{sec:hoeffding}) and Chernoff trigger (\abvSec\ref{sec:chernoff}).
			Over each red interval, the quadratic cost \eqref{eq:cost} is computed yielding the sampled $\hat{J}_N(j)$, while the data in-between remains unused.
			Hence, the Hoeffding discards a significant part of the collected data, in order to ensure approximate independence between samples of $J_N$.
			In contrast to this, the Chernoff trigger uses all data points by taking a single cost sample over a longer horizon.
		}
		\label{fig:triggerinterval}
	\end{figure}

	Thus, we ensure approximately independent samples by waiting for $r$ data points between each $N$ data points long cost sample (\cf \abvFig\ref{fig:triggerinterval}).
	The horizon of the last cost-sample ends at the current time step, which allows the trigger to take the most recent data into account.
	Therefore, we obtain the summation set
	\begin{equation}
		\mathfrak{\hoeffdingL}(k) \coloneqq \left\lbrace k-(N+r)i \middle|i\in 0,\dotsc,\hoeffdingL-1\right\rbrace \label{eq:hoeffding:summation_set}
	\end{equation}
	for the sample average in \eqref{eq:hoeffding:trigger}, which by construction has cardinality $\hoeffdingL$.
	Considering the definition of the cost \eqref{eq:cost}, the indices in $\mathfrak{\hoeffdingL}(k)$ mark the end of each red interval in \abvFig\ref{fig:triggerinterval}.
	In total, the trigger needs the last ${\hoeffdingL}(N+r)-r$ states as data, of which it only uses ${\hoeffdingL}N$ for approximating the mean.

	Strictly speaking, approximate independence still does not allow us to apply Hoeffding (\abvThm\ref{thm:hoeffding:inequality}). Therefore, for the time being, we assume the approximation is exact and then apply Hoeffding to obtain the thresholds. The previously introduced Chernoff trigger (\abvSec\ref{sec:chernoff}) solved this issue in an elegant and clean way by incorporating the correlations directly into the trigger via the MGF.
    
	\begin{thm}[Hoeffding Trigger]\label{thm:hoeffding:trigger}
		Let Assumptions~\ref{asm:stability} and \ref{asm:convergence} hold, assume the conditions of Lemma~\ref{lem:hoeffding:bound} are satisfied, and the samples $J_N(k-(N+r)i), i=0, \dotsc, \hoeffdingL-1$ are mutually independent.
		Further, let $\eta$ denote the desired confidence level and $\kappa$ is chosen as
		\begin{subequations}\begin{align}
			\kappa &= \sup_j \left\lbrace J_{N}(j)\right\rbrace \sqrt{-\frac{{\hoeffdingL}}{2} \ln \frac{\eta}{2}} \\
			&= \alpha^2 N \lambda_{\max}\!\left(W^\T Q W\right) \sqrt{-\frac{{\hoeffdingL}}{2} \ln \frac{\eta}{2}}\,.
		\end{align}\end{subequations}
		Then, the probability of
		triggering with \eqref{eq:hoeffding:trigger}, while the model coincides with the ground truth, is bounded by
		\begin{equation}
			\proba\!\left[\left\lvert \sum_{i=0}^{{\hoeffdingL}-1} J_{N}(k-(N+r)i)  - {\hoeffdingL} \expectation\!\left[J_{N}\right]\right\rvert\ge \kappa\right]\! \le \eta.
		\end{equation}
	\end{thm}
	\begin{IEEEproof}
		By construction, we can apply Hoeffding's inequality (\abvThm\ref{thm:hoeffding:inequality}) to $J_N$ at the sampling instances $\mathfrak{\hoeffdingL}(k)$.
		The bound is given by Lemma~\ref{lem:hoeffding:bound} as $b_i\equiv0$ and $a_i\equiv\sup J_N$.
		As the same inequality can also be applied to $-J_N$, we get the combined inequality
		\begin{align*}
		&\proba\!\left[\left\lvert\sum_{i=0}^{{\hoeffdingL}-1} J_N(k-(N+r)i)-{\hoeffdingL}\mu\right\rvert\ge \kappa\right] \\
		&\le 2\exp\left(\frac{-2\kappa^2}{{\hoeffdingL}\left(\sup J_N\right)^2}\right) = \eta.
		\end{align*}
		We set $\eta$ to coincide with the upper bound and rearrange for $\kappa$. Thus, we obtain
		\begin{align*}
		\eta&=2\exp\left(\frac{-2\kappa^2}{{\hoeffdingL}\left(\sup J_N\right)^2}\right)\\
		\kappa^2&=-\left(\sup J_N\right)^2\tfrac{{\hoeffdingL}}{2}\ln\tfrac{\eta}{2}.
		\end{align*}
		Then, the result is obtained by taking the square root and inserting the value for $\sup J_N$ from the Lemma~\ref{lem:hoeffding:bound}.
	\end{IEEEproof}

	In practice, $r_0$ is chosen large enough so that Lemma~\ref{lem:hoeffding:independence} ensures approximate independence of the samples $J_N(k-(N+r)i)$.
	Thus, we analyze if the empirical mean actually converges to the analytically derived expected value, while the technical details ensure that we avoid distorting the cost with random short term effects.

    \begin{rem}
        Following the same principles, it is possible to derive alternative triggers that consider different error terms or higher moments.
        However, we did not observe any improved performance of such triggers compared to the mean-based Hoeffding trigger.
        Considering relative instead of absolute errors is also possible, however, it does not improve the triggering behavior.
        We confirmed this in numerical investigations (analogous to \abvSec\ref{sec:numsim}) for a variance-based Hoeffding trigger, with similar theoretical guarantees, and showed that it yields no significant advantage over the mean trigger. 
        The Chernoff trigger does not suffer from the same limitations as the Hoeffding-based design.
    \end{rem}
	
	\section{Numerical Simulation}\label{sec:numsim}
	Next, we will numerically study the trigger architecture, as shown in \abvFig\ref{fig:architecture}.
	We will illustrate the triggering behavior and, in particular, investigate how well model change is detected with each trigger.

	\subsection{Setup} \label{sec:numsim:setup}
	Initially, a 5 dimensional system $(A_\mathrm{o},B,V)$ is randomly generated, by sampling the matrices $A_\mathrm{o}-\identity\in\mathbb{R}^{5\times 5}$, $B\in\mathbb{R}^{5\times 1}$, and $\sqrt{V}\in\mathbb{R}^{5\times 5}$
	elementwise from a uniform distribution between $\pm1$.
	The initial state is sampled from the asymptotic distribution of the closed-loop system, in order to fulfill Assumption~\ref{asm:convergence}.

	Next, we introduce the model $(\tilde{A}_\mathrm{o},\tilde{B},\tilde{V})$, which is used to compute the feedback controller and to derive the triggering thresholds. Initially, we set the model to the exact system parameters in order to demonstrate that the cost behaves as expected. Later on, we will distort the system dynamics $(A_\mathrm{o},B,V)$ to create a gap between model and true system parameters. For the model-based controller, we use LQR state feedback with unity weight matrices.

	The system is simulated for $50\,000$ time steps.
	At each time step, the cost and trigger value is computed as described below for each trigger.
	If the trigger detects a system change, then the model is set to the true parameters, \ie $(\tilde{A}_\mathrm{o}, \tilde{B}, \tilde{V}) \leftarrow ({A}_\mathrm{o}, {B}, {V})$.
	Thus, we abstract for the time being the actual model learning to setting the model parameters to the true values.
	While this, of course, is not possible in reality, for the simulation we are for now mainly interested in the behavior of the trigger.
	The learning part will be considered later in \abvSec\ref{sec:chernoff:hardware}.

	In order to simulate system changes, which the trigger should detect, we alter the system every $10\,000$ time steps without adjusting the model, trigger, nor controller.
	First, we tried sampling the new system dynamics $(A'_o,B',V')$ exactly the same way as for the initial system. However, these changes are usually quite significant and easy to detect.
	Thus, we bound the change with the aid of an additive model increment
	\begin{equation}
		\Delta = \beta\frac{(A_\mathrm{o}',B',V')-(A_\mathrm{o},B,V)}{\left\lVert(A_\mathrm{o}',B',V')-(A_\mathrm{o},B,V)\right\rVert_2},
	\end{equation}
	where $\beta\in\left(-0.1,0.1\right)$ is also sampled from an uniform distribution.
	Thus, the new system is obtained by adding $\Delta$ to the old system.
	If the resulting system is uncontrollable, a different increment is generated by sampling again.
	We do not enforce stability after altering the system since any threshold $\kappa^+$ will be reached eventually, and thus, triggering is trivial when the system is unstable.

	\paragraph{The Chernoff trigger} is computed from the model as described in \eqref{eq:chernoff:trigger} with a horizon of $N=200$ and $\eta=1\%$.
	A system change is detected, when the trigger value $\psi=J_N(\cdot)$ leaves the interval $(\kappa^-,\kappa^+)$.
	With the resulting model update, we have to recompute the trigger thresholds $\kappa^\pm$.

	\paragraph{The Hoeffding trigger} uses the simpler design \eqref{eq:hoeffding:trigger} with the increased misfire probability $\eta=25\%$.
	For the moving average sampling we use $N=r=60$ and ${\hoeffdingL}=20$ (\cf \abvFig\ref{fig:triggerinterval}).
	The adjustment of $N$ and $\eta$ compared to the Chernoff trigger are required for detection with this trigger due to the significant conservatism and theoretical shortcomings of the design. 
	The required state-bound $W$ is set to the covariance of the state and $\alpha=18$.
	These bounds are constant throughout the simulation.
	Thus, the threshold $\kappa$ remains constant as well, while the mean $\expectation[J_N]$ is the only part of the trigger that changes with model updates.

	For both triggers, we use the same random seed allowing for direct comparison of the result as shown in \abvFig\ref{fig:chernoff:numsim} and \abvFig\ref{fig:hoeffding:numsim}.

	\subsection{Results}

	\begin{figure}
		\centering
		\includegraphics[width=\linewidth]{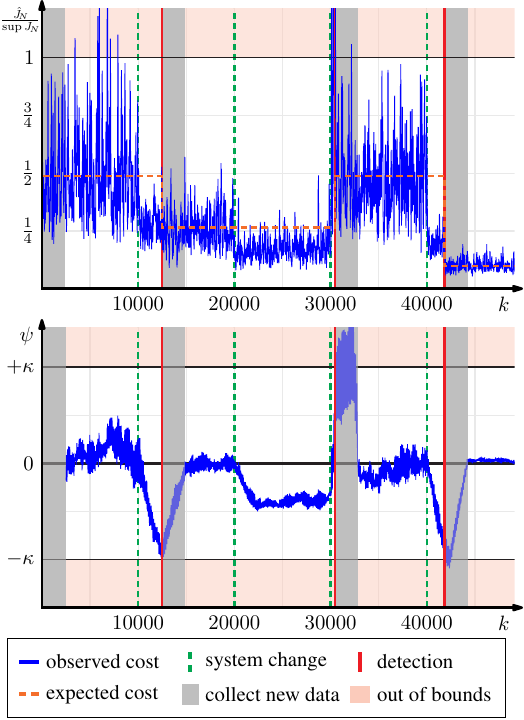}
		\caption{
			Numerical simulation of the Hoeffding trigger on a randomly generated 5-dimensional system.
			In the lower graph the trigger statistic $\psi$
			is shown (blue line).
			Every $10\,000$ time steps (green lines), the system is randomly altered in order to simulate change.
			Leaving the confidence bounds $(-\kappa, \kappa)$ triggers learning (red line). Then, the model is set to the true system parameters, a new feedback gain computed, and a new value for $\expectation[J_N]$ derived.
			In the upper graph, the normalized cost is shown (in blue) and the model-based expected value $\expectation\bigl[J_N\bigr]$ (orange line).
		}
		\label{fig:hoeffding:numsim}
	\end{figure}
        
	\begin{figure}
		\centering
		\includegraphics[width=\linewidth]{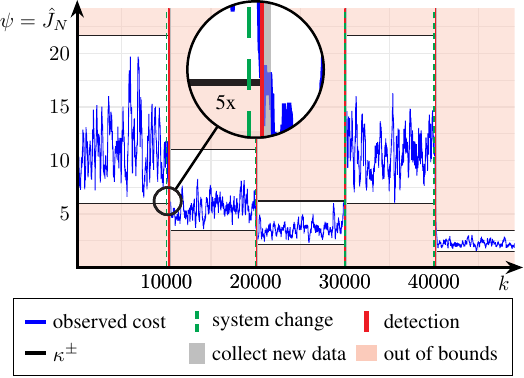}
		\caption{%
			Numerical simulation of the Chernoff trigger on a 5-dimensional system with random $A_\mathrm{o}$ and $B$ matrices.
			At the indicated time step (green), the entries of the $A$ and $B$-matrices are randomly altered in order to simulate a change in the dynamics.
			This change is detected at the red lines, at which point the model, the feedback controller, and the thresholds $\kappa^\pm$ are updated.
		}
		\label{fig:chernoff:numsim}
	\end{figure}
    
    Next, we look at the numerical performance for both triggers.
    The shown roll-out was pick as it displays many interesting effects observed throughout various roll-outs and is a good representation of the observed behavior.
    In this run neither instability, significant violation of the upper bound on the cost for the Hoeffding trigger, nor the issues that arise from the inactive trigger after a detection are shown.
    Their effects are obvious and observed in other roll-outs.
    
    In the upper plot of \abvFig\ref{fig:hoeffding:numsim}, we see the normalized cost, and in the lower one, the trigger statistic $\psi$ for the Hoeffding trigger.
    The Chernoff trigger requires only a single plot in \abvFig\ref{fig:chernoff:numsim}, since it utilizes the cost $\hat{J}_N$ directly as trigger statistic $\psi$.
	The system is distorted every $10\,000$ time steps (green dashed line).
	A detection (red line) occurs when the trigger value $\psi$ (in blue) hits either threshold $\kappa$, which can be seen in the (lower) plot for trigger statistic.
    Recall that each trigger uses a different horizon for the cost, hence due to the longer horizon, the cost for the Chernoff trigger looks smoother.
    
    Foremost, this example illustrates the shortcomings of the mean-based design in contrast to the normal operation of the Chernoff trigger.
    In the following, we consider the individual system changes and their detection in detail. 
    
   	At $k=10\,000$, the first system change occurs, which is detected after $2\,480$ steps with the Hoeffding trigger.
   	This is a significant delay between changing the dynamics and detecting said change.
   	However, these transient effects are not surprising due to the large amount of data required by the trigger with its window of length ${\hoeffdingL}(N+r)=2400$.
   	Thus, it might take some time (\cf \abvFig\ref{fig:triggerinterval}) until the new dynamics affect the entire horizon of the moving average.
   	This effect can also be observed at $k=40\,000$, where the detection takes $1\,838$ steps.
    The Chernoff trigger, in contrast, does not suffer from this issue.
    While the achieved detection times of between $50$ and $280$ steps in this simulation are only possible due to the better bounds, of course, the short window of states considered also expedites the detection.

   	The Hoeffding trigger does not detect the change at $k=20\,000$, which demonstrates the downsides of this trigger design.
   	The upper bound is too conservative, and thus, the change cannot be detected with the given amount of data.
   	Considering more data would be possible, however, it would also increase the delays even further.
   	Smaller upper bounds are not possible since these were designed for the initial system by hand.
   	In principle, it would be possible to change the bounds during triggering, however, it is not trivial to do so automatically.
    The Chernoff is designed to deal with possible unbounded cost, insofar the order of magnitude of the cost has no impact on the triggering performance.

   	At $k=30\,000$, there is a fast increase of the cost and a lot of deviation within the signal.
   	This allows even the Hoeffding trigger a fast detection after only $435$ steps.
   	Yet, the Chernoff trigger is still significantly faster in detecting the change within $50$ steps.
    
    Secondly, for the Chernoff trigger, we can see that the bounds are tight in the sense that the cost stays within the confidence interval, but also comes close to the edges. Thus, the probability mass is distributed as intended. Even though the trigger uses little data, it rarely misfires -- not once in this run. In the next section, we will present a large scale experiment to further investigate false positives and trigger delays. We obtain a misfire rate of less than 0.01\%\ over the simulated four billion time steps, which is, as designed, less than $\eta = 1\%$.

    \subsection{Discussion}
	\subsubsection{Hoeffding Trigger}
	First of all, the Hoeffding trigger does a decent job at detecting change.  However, there are some downsides and limitations.

	The large amount of data required by the trigger affects detection.
	Clearly, there is a significant delay in the detection, which corresponds to the magnitude of the time window.
	Further, it prevents the detection of quick changes since new data has to be gathered after each model update.

	Furthermore, bounding the cost derived from Lemma~\ref{lem:hoeffding:bound} is an issue. In order to apply Hoeffding's inequality rigorously, we need the bound to hold everywhere. However, this has a significant impact on the detection rate since the possibility of high costs increases the possible confidence interval.
	Additionally, Hoeffding's inequality is per se based upon the worst-case distribution and, thus, not very tight.

	Nonetheless, there is sufficient information in the cost signal to detect changes in the system dynamics reliably.
	Yet, this trigger only exploits a small part of the available data and, hence, it only achieves sub-optimal detection times and misses some changes.

	\subsubsection{Chernoff Trigger}
	Since the trigger thresholds are tailored to the actual distribution, we can see a  superior performance.
	In particular, the adaptivity of the thresholds to different magnitudes of process noise can be clearly seen in \abvFig \ref{fig:chernoff:numsim}. For instance, at $k =\text{40\,000}$, and afterward, there is little deviation in the cost, and this is also captured in the bounds. However, between $k=0$ and $k = \text{10\,000}$ there are strong oscillations in the signal. Nonetheless, the interval fits nicely.

	Furthermore, the shorter time-window of only $200$ instead of $2\,400$ steps results in faster and more reactive triggering.
	However, therein lies a trade-off with random fluctuations and unmodelled disturbances. These can have large impacts on the trigger value $\psi$, as they are not averaged out. Therefore, we will discuss further possibilities to robustify the trigger in \abvSec\ref{sec:chernoff:hardware}, where we consider a hardware experiment.

	\subsection{Detection Delay}
	To study the detection delays of this trigger (\ie the time between changing the system and the trigger detecting the change), we ran large scale Monte Carlo simulations using the same setup as before.
	However, we ignored unstable changes and resampled when this happened.
	Each roll-out was simulated an hour of wall-time before a new roll-out with a different random seed was started.
	The restarts are required as the used pseudo-random number generator for the noise and system changes has only limited entropy.
	Eight roll-outs were computed in parallel on an Intel\textsuperscript{\textregistered} Xeon\textsuperscript{\textregistered} W-2123 3.6\,GHz 8-core processor, for a total of two weeks accumulating a total of 3\,976\,360\,000 simulated time steps with 397\,636 system changes.

	\subsubsection{A System Change Metric}
	Our hypothesis is that the detection delay depends on the \emph{size} of the system change.
	Thus, we require a metric to quantify this.
	For this purpose, we compare the system norm before and after the change.
	Considering the stochastic nature of the problem, using the $H_2$-norm seemed suitable. This norm is closely linked to the steady-state covariance of the system  when driven by white noise input.
	In detail, the $H_2$-norm for an input-output system $\mathcal{G}$ with input $w$ and output $y$ is defined as
	\begin{equation}
		\begin{gathered}
			\bigl\lVert \mathcal{G} \bigr\rVert_{H_2} = \sqrt{\lim_{k\rightarrow \infty}\trace\expectation\bigl[\,y_k^\T y_k\,\bigr]},
			\\
			\expectation\left[w_i\right]=0\,,\quad\expectation\bigl[\,w_i w_j^\T\,\bigr]=\delta_{ij}\identity\,.
		\end{gathered}
	\end{equation}
	Further, we decided to measure the system change as
	\begin{equation}
		\delta_{\mathrm{sys}}\coloneqq \frac{%
			\left\lVert \left\lbrack\begin{array}{c|c}
			A_{\mathrm{new}} & \!\sqrt{V_{\mathrm{new}}}\!\\
			\hline
			\identity & 0 \\
			\end{array}\right\rbrack \right\rVert_{H_2}
		}{%
			\left\lVert \left\lbrack\begin{array}{c|c}
			A_{\mathrlap{\mathrm{old}}\phantom{\mathrm{new}}} & \!\sqrt{V_{\mathrlap{\mathrm{old}}\phantom{\mathrm{new}}}}\!\\
			\hline
			\identity & 0 \\
			\end{array}\right\rbrack \right\rVert_{H_2}%
		},
	\end{equation}
	with the old and new closed-loop system matrix $A$ and the square root of process noise covariance $V$.
	The $\left\lbrack {\mathrlap{-\!\!-} \hphantom{-\!} \mathclap{\mid} \hphantom{\!-}} \right\rbrack$-notation represents the system and is commonly used in the robust control community~\cite[\abvSec3]{zhou1998robust}.
	We use the square root of $V$ as input since this will transform the white noise of the norm into the actual Gaussian noise.

	\subsubsection{Estimating Probability Density $\proba\bigl(T_D\big|\delta_{sys}\bigr)$}
	Additionally, we need to clarify how we measure the detection delay $T_D$.
	We define the delay as the number of time steps from the system change, which is instantaneous in the simulation, to the first time step, the trigger threshold is passed.
	In particular, we do not consider detection only after the threshold is passed for some time, as implemented on hardware (\cf \abvSec\ref{sec:chernoff:hardware}).

	The Monte Carlo simulation yields samples from the joint probability $\proba\bigl(T_D,\delta_{\mathrm{sys}}\bigr)$ of the detection delay $T_D$ and system change $\delta_{\mathrm{sys}}$.
	From these samples, we compute an estimate for the probability density function using a Gaussian kernel smoothing with the \textsc{Matlab}\textsuperscript{\textregistered}-command\footnote{ \textsc{Matlab}\textsuperscript{\textregistered}, \textsc{Simulink}\textsuperscript{\textregistered} are registered trademarks of The MathWorks, Inc.\label{foot:matlab}} \texttt{ksdensity}\footnote{https://www.mathworks.com/help/stats/ksdensity.html} on $\bigl(T_D,\log_{10}\delta_{\mathrm{sys}}\bigr)$.
	Applying the logarithm is beneficial from a numerical point of view.

	Clearly, the system changes $\delta_{\mathrm{sys}}$ are not uniformly distributed, and thus, the joint probability density is difficult to interpret.
	Hence, we condition on $\delta_{\mathrm{sys}}$ to obtain the conditional probability $\proba\bigl(T_D\big|\delta_{\mathrm{sys}}\bigr)$.
	In order to do that we need to compute the density function for the marginal probability $\proba\bigl(\delta_{\mathrm{sys}}\bigr)$, for which we again apply a Gaussian kernel smoothing with \texttt{ksdensity} on the
	logarithm of $\delta_{\mathrm{sys}}$.
	The conditional probability is then computed by division.

	\subsubsection{Results}
	\begin{figure}
		\centering
		\includegraphics[width=\linewidth]{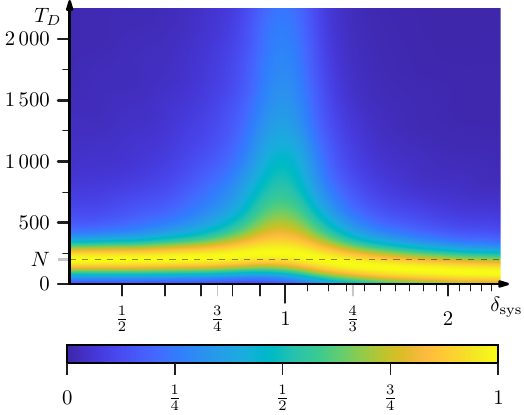}
		\caption{%
			The probability density estimate $\proba\bigl(T_D\big|\delta_{\mathrm{sys}}\bigr)$ of the detection delay $T_D$ of the Chernoff trigger conditioned on the relative change $\delta_{\mathrm{sys}}$.
			The estimate is obtained with a Gaussian smoothing kernel from a Monte Carlo simulation with $397\,636$ samples using the setup described in \abvSec\ref{sec:numsim:setup}.
			The maximal value of the density for any fixed $\delta_{\mathrm{sys}}$ is normalized to one, thus the most likely delay for a given system change $\delta_{\mathrm{sys}}$ is colored in yellow.
		}
		\label{fig:chernoff:delay}
	\end{figure}
	In \abvFig\ref{fig:chernoff:delay}, the obtained density function for $\proba\bigl(T_D\big|\delta_{\mathrm{sys}}\bigr)$ is shown.
	For the visualization the graph renormalized such that
	$\forall\vartheta : \max_\tau\bigl\lbrace\proba\bigl(T_D=\tau\,\big|\delta_{\mathrm{sys}}=\vartheta\bigr)\bigr\rbrace=1$.
	We can see that a change is most likely detected after $N$ time step, as we observed earlier when considering just a single roll-out.

	Since we are using a relative metric, a value of 1 implies that there was no change in the system.
	We can clearly see in \abvFig\ref{fig:chernoff:delay} that the probability mass is rather concentrated for significant changes in the system (\ie $\delta_{\mathrm{sys}} \ll 1$ and $\delta_{\mathrm{sys}} \gg 1$). Moving towards $\delta_{\mathrm{sys}} = 1$, we can observe that the detection time increases and also the variance. More and more probability mass is pushed towards large detection times. Exactly at $\delta_{\mathrm{sys}} = 1$, the triggering should be purely due to false positives.
	However, we did not record any data points exactly at $\delta_{\mathrm{sys}} = 1$ since this event has probability zero. Further, there are some smoothing effects in \abvFig\ref{fig:chernoff:delay}, in particular, around $\delta_{\mathrm{sys}} = 1$.

	\section{Hardware Experiment: Rotary Pendulum}\label{sec:chernoff:hardware}
	While the previous numerical examples showed the effectiveness of the proposed triggers, we now investigate their efficacy under real-world and, thus, non-ideal conditions.
	We consider the pole-balancing performance of a rotary pendulum. We focus on the Chernoff trigger, which proved to be superior in the numerical experiments and has better theoretical properties.

	\subsection{Experimental Setup}
	\begin{figure}
		\centering
		\includegraphics[width=\columnwidth,height=0.3\textheight,keepaspectratio]{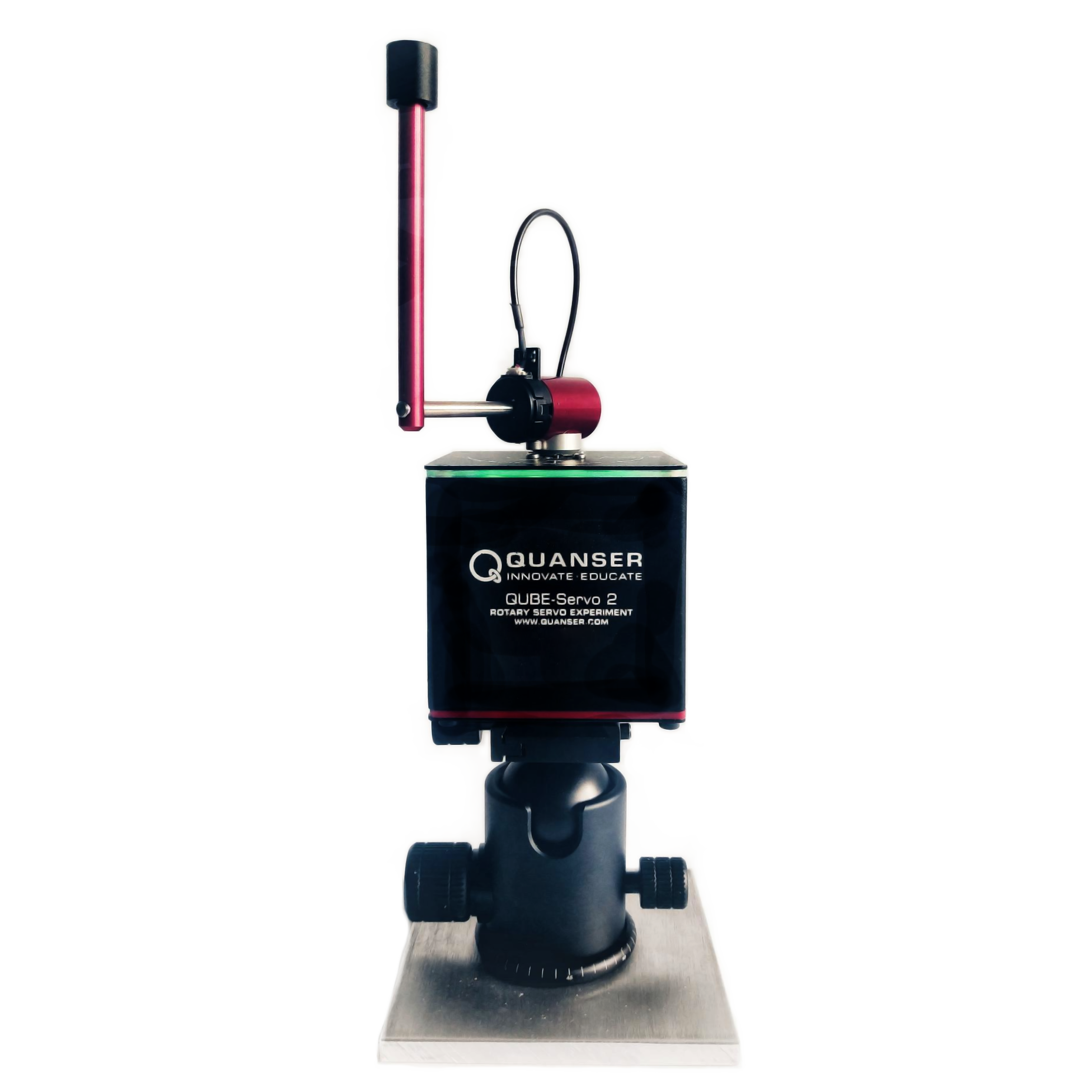}
		\caption[Hardware Setup]{The experiment setup consists of a Quanser QUBE Servo 2 rotary pendulum that is mounted on top of a tripod. Here, the plant is shown just after swing-up.}
		\label{fig:chernoff:hardware}
	\end{figure}
	We implemented the proposed learning trigger on a modified Quanser rotary pendulum~\cite{Quanser2016}, as shown in \abvFig\ref{fig:chernoff:hardware}.
	We denote the axial rotation as $\theta$ and the angle between the vertical position and the pendulum as $\alpha$.
	The setup allows us to directly measure both angles and the velocity in $\theta$.
	For the velocity in $\alpha$, we use the provided high-pass filter to approximately differentiate the angle.

	The pendulum has been modified in such a way that allows for changing the dynamics in two distinct ways.
	First, the base of the platform is mounted on top of a tripod.
	Using its ball joint, we can tilt the pendulum freely in any direction.
	Additionally, a magnet is attached to the top end of the pendulum, which allows us to change the inertia by adding magnetic weights.
	By varying these two, we can change the system dynamics and validate if the trigger is able to detect these.

	Using \textsc{Simulink}\textsuperscript{\textregistered}\textsuperscript{\ref{foot:matlab}}, we implemented a switched controller running at a sampled rate of 500\,Hz.
	An LQRI controller, \ie an LQR with an additional integrator, is used to stabilize the upright position.
	While outside the approximately linear region (\textpm\,20\textdegree), a nonlinear swing-up controller is used to bring the inverted pendulum back to the linear region.
	Including the artificial integrator on $\theta$ as an extra state $e \coloneqq \int\theta\,\mathrm{d}t$ in the plant model yields a five dimensional system with the state $x^\T=\bigl[\theta,\alpha,\dot{\theta},\dot{\alpha},e\bigr]$.

	\subsection{Event-triggered Learning Design}
	On this linear controller, we apply the proposed event-triggered learning strategy (\ie the Chernoff trigger \eqref{eq:chernoff:trigger}) as shown in \abvFig\ref{fig:architecture}.
	The empirical cost $\hat{J}_N(k)$ is computed at every time step with a horizon of $N=\text{200}$.
	In order to avoid detecting an instantaneous disturbance, like for example, the jerk introduced by enacting a system change via tilting or weights, we modify the trigger slightly.
	We introduce the additional condition that the threshold has to stay surpassed for more than 10 seconds.
	Thus, we achieve more robust detection against strong short term disturbances. 

	When the trigger detects a change, a learning experiment is started.
	For this, the trigger and integrator are disabled, as they would react to the learning excitation.
	However, this introduces an initial disturbance, which we overcome by waiting a few seconds until the system returns to steady-state.

	For the learning experiment, an artificial excitation signal is added to the control input. Choosing a signal that is both sufficiently exciting~\cite{Ljung2009} for possibly changed dynamics of the pendulum and avoids the hardware constraints in $\theta$ turned out to be a nontrivial problem on this experiment. In general, it is difficult to design well-behaved excitation signals a priori. Here, we apply a carefully tuned chirp signal, which first increasing and then decreasing frequency.

	Learning itself is performed using prediction error minimization from the \textsc{Matlab}\textsuperscript{\textregistered} System Identification toolbox\textsuperscript{\ref{foot:matlab},}\footnote{https://de.mathworks.com/help/ident/ref/pem.html}, with an initial guess based on least square estimation.

	Due to the nonlinear, state-dependent, non-white noise of the actual pendulum, we cannot use the data to estimate the process noise directly.
	Instead, we record a few seconds of steady-state behavior with the new controller, including the integrator, and estimate the covariance.
	Thereby, we obtain a linear Gaussian approximation of the process noise around the steady-state.

	With the linear model and process noise, we can compute new trigger thresholds $\kappa^\pm$ (\cf \abvThm\ref{thm:chernoff:trigger} and Equation \eqref{eq:kappachernoff}), which completes the model update.
	
	\subsection{Results}
	\begin{figure}
		\centering
		\includegraphics[width=\linewidth]{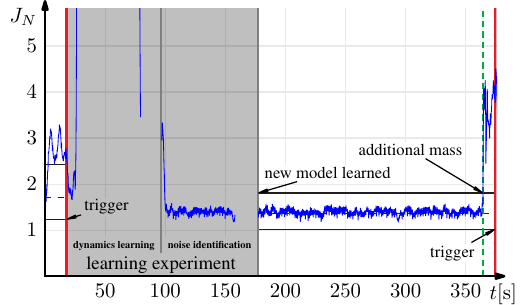}
		\caption{%
			Experimental run of the Chernoff trigger on the rotary pendulum (\abvFig\ref{fig:chernoff:hardware}).
			The black lines indicate $\kappa^+$ and $\kappa^-$, respectively.
			Additionally, the model-derived expected value (\cf Lemma~\ref{lem:mgf:moments}) of the cost is shown as a dashed line.
			At the red lines, a change is detected, and thus, a learning experiment triggered.
			During this model update, the trigger is offline, as indicated in grey.
			At the green dashed line, the physical system is changed by adding a weight.
		}
		\label{fig:chernoff:rotpen:run1}
	\end{figure}

	In \abvFig\ref{fig:chernoff:rotpen:run1}, we can see the (measured) cost of an exemplary run of the Chernoff trigger on the hardware setup.
	The setup has been initialized with a sightly incorrect model.
	That is, the parameters of the first-principle model, provided by Quanser, have been changed slightly.
	Both the initial controller and the initial bounds of the trigger have been computed based on this faulty model.
	The main goal is to show that we are able to detect change systematically and, thus, effectively reduce the cost by updating the controller.

	As we can see at the very beginning of \abvFig\ref{fig:chernoff:rotpen:run1}, the measured cost does not lie within the interval $(\kappa^-,\kappa^+)$.
	Since we used an inaccurate model to design the feedback controller, this is to be expected.
	The cost quickly rises above the threshold, however, at $7.758\,\mathrm{s}$, it has a down crossing caused by random effects.
	Hence, the change is only detected after $17.758\,\mathrm{s}$.

	After the model has been updated (end of the learning experiment), we see that the cost lies within the new trigger interval.
	Indeed, it oscillates nicely around the computed expected value.

	Most importantly, we effectively reduce the cost. Before we triggered learning, the cost signal was significantly higher (roughly two times on average) than after updating the model and controller. Further, we also obtain new thresholds to detect an additional change in the dynamics.

	After approximately six minutes, we add a weight to the pendulum, which is indicated by the green line in \abvFig\ref{fig:chernoff:rotpen:run1}.
	At $374.272\,\mathrm{s}$, the trigger detects this change.
	We want to emphasize again that this detection is not due to the initial disturbance.
	Instead, it is due to the change in dynamics and, thus, a different cost distribution, which we successfully detect.

	\subsection{Hardware and Implementation}
	Around $80\,\mathrm{s}$ and $160\,\mathrm{s}$, there are two chunks of missing data, which is an artifact of our implementation. During these times, the updates of the system matrices and trigger thresholds $\kappa^\pm$ were computed. During these computations, no data was collected. However, this did not influence the controller and, thus, the presented results.

	The required connection wire between the rotating sensor for measuring $\alpha$ and the base introduces a time-variant nonlinearity into the setup.
	As this wire randomly twists during operation and swing-up, the equilibrium state may change in $\theta$.
	Additionally, the wire applies a force towards some $\theta$, which may not be zero.
	While these effects have little impact on the controller, they pose a problem for linear system identification and especially the noise estimation.
	Thus, we can only consider runs, where the wire remains close to its correct state.

	Also, tilting the pendulum in any direction by at least one degree yields an interesting problem.
	Detecting the change is straightforward with our approach.
	However, the new system is at least affine and does not have an upper equilibrium without input.
	Thus, it is challenging to identify new bounds for the trigger.
	Handling such a system might be possible but requires some adjustment in our approach.

	\section{Conclusion}\label{sec:conclusion}
	In this article, we propose a Chernoff type learning trigger to trigger model learning when needed based on the distribution of LQR cost function.
	Thus, we obtain a highly flexible control scheme that leverages well-known results from LQR and combines them with tools from statistical learning theory.
	By explicitly computing the moment generating function of the LQR cost function, we are able to tailor learning triggers tightly to the problem at hand.

	The derived learning triggers are extensively validated in numerical simulation and yield the expected results.
	Furthermore, we show in a hardware experiment that the approach can be applied to a real system, where it effectively detects the change and reduces the incurred control cost and steady-state variance.

	\section*{Acknowledgement}\label{sec:acknowledgement}
	The authors thank Jonathan Fiene and Ruben Werbke for their support with the experimental setup.

	\bibliographystyle{IEEEtranDOI}
	\bibliography{IEEEabrv,database.bib}

	\ifCLASSOPTIONpeerreview\else
	\begin{IEEEbiography}%
		[{\includegraphics[width=1in,height=1.25in,clip,keepaspectratio]{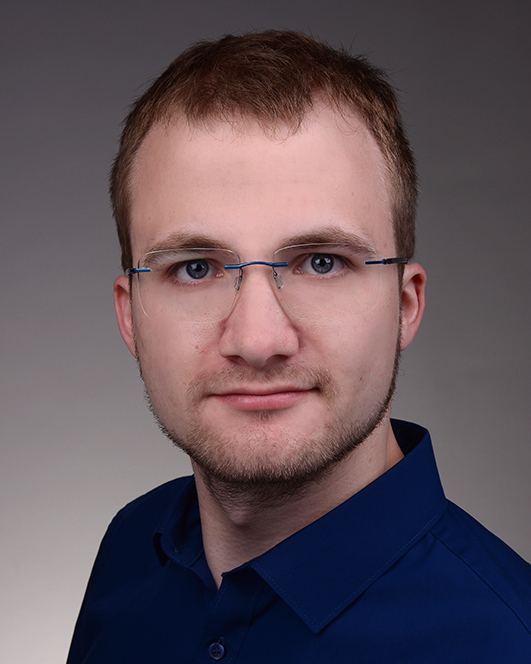}}]%
		{Henning Schl\"uter}%
		received the B.\,Sc.\ and M.\,Sc.\ degrees in engineering cybernetics from the University of Stuttgart in Germany, in 2017 and 2019, respectively.
		He has since been a doctoral student at the Institute for Systems Theory and Automatic Control of the University of Stuttgart under the supervision of Prof.~Frank Allg\"ower and a member of the International Max Planck Research School for Intelligent Systems.
		His research interests are in the area of model predictive control, and stochastic control.
	\end{IEEEbiography}
	\begin{IEEEbiography}%
		[{\includegraphics[width=1in,height=1.25in,clip,keepaspectratio]{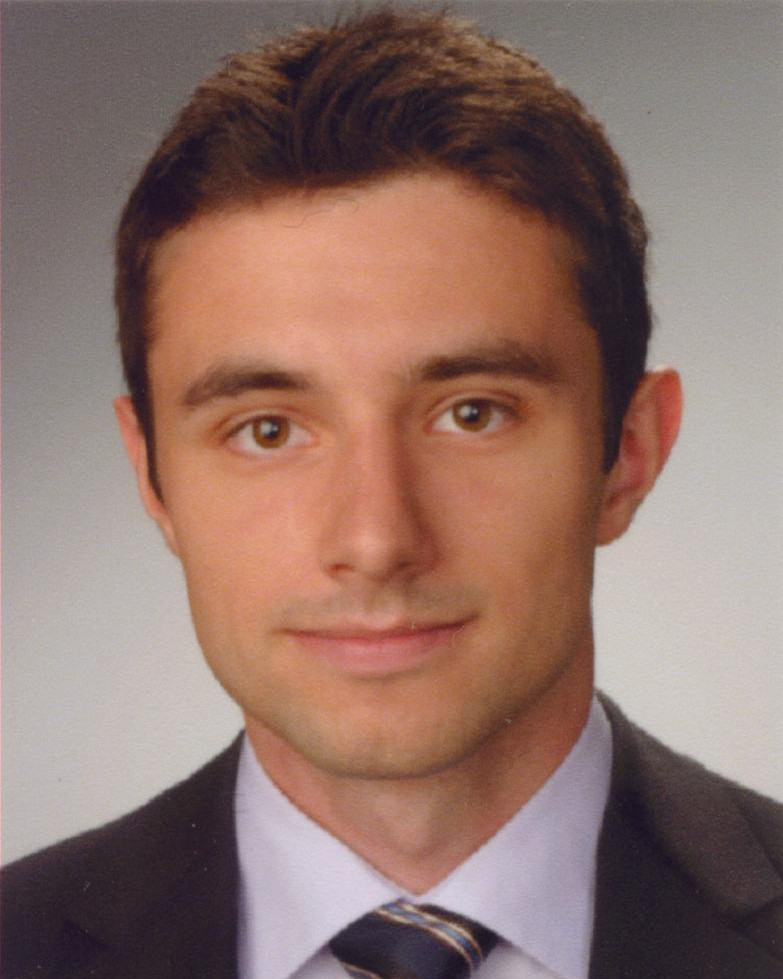}}]%
		{Friedrich Solowjow}%
		received B.\,Sc.\ degrees in Mathematics and Economics from the University of Bonn in 2014 and 2015, respectively, and a M.\,Sc.\ degree in Mathematics also from the University of Bonn in 2017. He is currently a Ph.\,D.\ student in the Intelligent Control Systems Group at the Max Planck Institute for Intelligent Systems, Stuttgart, Germany and a member of the International Max Planck Research School for Intelligent Systems.
		His main research interests are in systems and control theory and machine learning.
	\end{IEEEbiography}
	\begin{IEEEbiography}%
		[{\includegraphics[width=1in,height=1.25in,clip,keepaspectratio]{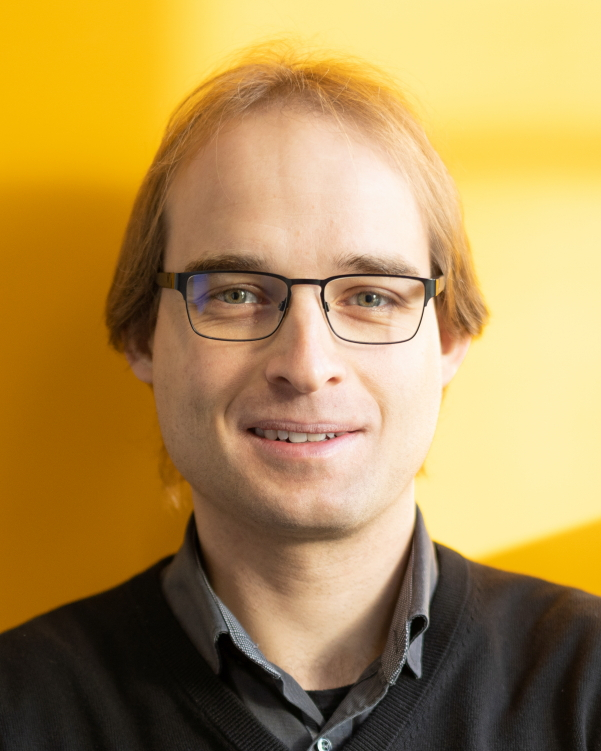}}]%
		{Sebastian Trimpe}%
		(M'12) received the B.\,Sc.\ degree in general engineering science and the M.\,Sc.\ degree (Dipl.-Ing.) in electrical engineering from Hamburg University of Technology, Hamburg, Germany, in 2005 and 2007, respectively, and the Ph.\,D.\ degree (Dr.\,sc.) in mechanical engineering from ETH Zurich, Zurich, Switzerland, in 2013.  Since 2020, he is a full professor at RWTH Aachen University, Germany, where he heads the Institute for Data Science in Mechanical Engineering.  Before, he was an independent Research Group Leader at the Max Planck Institute for Intelligent Systems in Stuttgart and Tübingen, Germany.  His main research interests are in systems and control theory, machine learning, networked and autonomous systems.  Dr.~Trimpe is recipient of several awards, among others, the triennial IFAC World Congress Interactive Paper Prize (2011), the Klaus Tschira Award for achievements in public understanding of science (2014), and the Best Paper Award of the International Conference on Cyber-Physical Systems (2019).
	\end{IEEEbiography}
	\fi
\end{document}